# Nonadiabatic Dynamics with Constrained Nuclear-Electronic Orbital Theory

*Zhe Liu, Zehua Chen, and Yang Yang\**

Theoretical Chemistry Institute and Department of Chemistry

University of Wisconsin-Madison

1101 University Avenue, Madison, WI 53706, United States

E-mail: yyang222@wisc.edu

**Abstract**

Incorporating nuclear quantum effects into nonadiabatic dynamics remains a significant challenge. Herein we introduce new nonadiabatic dynamics approaches based on the recently developed constrained nuclear-electronic orbital (CNEO) theory. The CNEO approach integrates nuclear quantum effects, particularly quantum nuclear delocalization effects, into effective potential energy surfaces. When combined with Ehrenfest dynamics and surface hopping, it efficiently captures both nonadiabaticity and quantum nuclear delocalization effects. We apply these new approaches to a one-dimensional proton-coupled electron transfer model and find that they outperform conventional Ehrenfest dynamics and surface hopping in predicting proton transfer dynamics and proton transmission probabilities, especially in the low momentum regime.

Many chemical and biological processes involve nonadiabatic transitions between electronic states, such as photoinduced charge and energy transfer in molecular systems,[1–3] proton-coupled electron transfer (PCET) in enzymatic and electrochemical reactions,[4–7] and electron transfer at metal-molecule interfaces in catalysis and surface chemistry.[7–9] To simulate these nonadiabatic processes, where the conventional Born-Oppenheimer approximation breaks down, many methods have been developed, including Ehrenfest dynamics[10], surface hopping,[10,11] the mixed quantum-classical Liouville equation,[12,13] semiclassical approaches,[14,15] and exact-factorization-based approaches.[16,17] Compared to the most accurate quantum dynamics methods,[18–20] these methods offer significantly reduced computational costs while maintaining reasonable accuracy. In practice, mixed-quantum-classical (MQC) methods, particularly surface hopping, have been widely used and successfully captured nonadiabatic effects in various model and real systems.[8,10,11,21]

Despite their success, the two popular MQC methods, Ehrenfest dynamics and surface hopping, treat nuclei classically and therefore cannot capture nuclear quantum effects (NQEs) such as zero-point energy (ZPE) and nuclear tunneling. In processes where NQEs play a crucial role, such as photoinduced hydrogen tunneling[22] and photoinduced PCET reactions,[4,5,23] developing nonadiabatic MQC methods that can capture NQEs is essential.

To incorporate NQEs into nonadiabatic dynamics, several efforts have been made to combine MQC methods with approaches capable of capturing NQEs, particularly path-integral-based approaches. For example, nonadiabatic isomorphic Hamiltonian dynamics have shown improved reaction rates at low temperatures compared to surface hopping results.[24] Nonadiabatic RPMD (NRPMD) has accurately simulated time-correlation functions,[25,26] and ring polymer surface hopping (RPSH)[25,27] has demonstrated higher accuracy than conventional fewest-switches surface hopping (FSSH), especially in the low initial momentum regime of the avoid crossing models where NQEs are more pronounced. Although these path-integral-based methods have so far been tested primarily on model systems with simple potential energy surfaces (PESs), they have provided compelling evidence that further theoretical development in this direction is important and promising. In addition to the path-integral-based methods, the combination of Ehrenfest dynamics and the nuclear-electronic orbital (NEO) framework has been developed and applied to real molecules,[28,29] which provided an alternative approach to simulate coupled nuclear-electronic dynamics.

Recently, our group developed the constrained nuclear-electronic orbital (CNEO) theory as an efficient way of incorporating NQEs, particularly quantum nuclear delocalization effects into molecular simulations.[30–32] CNEO belongs to the family of multicomponent quantum theories,[33–36] in which both electrons and nuclei are treated on the same footing with quantum mechanics. However, unlike traditional multicomponent approaches, the CNEO framework retains the conventional molecular geometry picture by imposing constraints on the expectation values of nuclear positions and setting them at classical molecular geometries.[37–45] By solving for the lowest energy state at each molecular geometry, the CNEO theory can yield effective PESs as functions of the nuclear expectation positions. Compared to conventional PESs, the CNEO effective PESs naturally include zero-point effects due to the delocalized nuclear picture. Past research has shown that with essentially the same computational cost, molecular dynamics on CNEO surfaces (CNEO-MD) significantly outperforms *ab initio* molecular dynamics (AIMD) in predicting hydrogen-related vibrational frequencies,[37–43] proton distributions,[44] reaction barriers,[45] and transition state rate constants.[45] However, past development and testing on CNEO have been limited to adiabatic processes on a single surface.

In this Letter, we introduce new methods for efficiently incorporating NQEs into nonadiabatic dynamics by combining the CNEO framework with conventional nonadiabatic dynamics methods, including Ehrenfest dynamics and surface hopping. We first derive the CNEO-Ehrenfest method, followed by the formulation of the CNEO fewest-switch surface hopping (CNEO-FSSH) algorithm. These two approaches are then applied to a one-dimensional proton-coupled electron transfer (PCET) model—the Shin-Metiu model[46]—and compared with the conventional Ehrenfest and FSSH methods, as well as quantum dynamics references. We find that, with essentially the same computational cost, these CNEO-based MQC approaches offer much improved accuracy over conventional MQC methods, especially in the low-momentum regime.

We begin by considering the total Hamiltonian of a molecular multicomponent quantum system, which consists of slow-moving particles, including quantum nuclei, and fast-moving particles, including all electrons. The total Hamiltonian can be written as

$$\begin{aligned}H(\mathbf{r},\mathbf{R}) &= T_\mathrm{n}(\mathbf{R}) + T_\mathrm{e}(\mathbf{r}) + V_\mathrm{ee}(\mathbf{r}) + V_\mathrm{en}(\mathbf{r},\mathbf{R}) + V_\mathrm{nn}(\mathbf{R}) + V_\mathrm{e}^\mathrm{ext}(\mathbf{r}) + V_\mathrm{n}^\mathrm{ext}(\mathbf{R}) \\ &\equiv T_\mathrm{n}(\mathbf{R}) + H_r(\mathbf{r},\mathbf{R})\end{aligned} \quad (1)$$

Here, **R** and **r** represent the nuclear and electronic coordinates, respectively. The terms in the first line correspond to the kinetic energy of the nuclei, the kinetic energy of the electrons, electron-electron interactions, electron-nucleus interactions, nucleus-nucleus interactions, the external potential for electrons, and the external potential for nuclei, respectively. In the second line, we group all terms except the nuclear kinetic energy into $H_r(\mathbf{r}, \mathbf{R})$, which represents the effective Hamiltonian for the electronic subsystem.

In principle, the exact multicomponent wave function can be obtained by solving the time-dependent Schrödinger equation using the full Hamiltonian. However, the exact quantum dynamical solution is often computationally intractable. To simplify the problem, we take the mean-field assumption and assume that the time-dependent multicomponent wave function can always be expressed as:

$$\Psi(\mathbf{r}, \mathbf{R}, t) = \phi(\mathbf{r}, t)\Omega(\mathbf{R}, t)e^{\frac{i}{\hbar}\int E_r(t')dt'} \equiv \phi(\mathbf{r}, t)\Omega(\mathbf{R}, t)e^{\frac{i}{\hbar}\varphi(t)} \quad (2)$$

Here $\phi(\mathbf{r}, t)$ is the wave function of the fast-moving electrons, while $\Omega(\mathbf{R}, t)$ describes the slow-moving nuclei. Following Tully's convention,[10] a phase factor has been extracted from the wave function with $E_r(t)$ defined as

$$E_r(t) = \int \Omega^*(\mathbf{R}, t)\phi^*(\mathbf{r}, t)H_r(\mathbf{r}, \mathbf{R})\phi(\mathbf{r}, t)\Omega(\mathbf{R}, t)d\mathbf{r}d\mathbf{R} \quad (3)$$

and the time evolution equations for the nuclear and electronic parts, respectively, are

$$i\hbar\frac{\partial}{\partial t}\Omega(\mathbf{R}, t) = T_n\Omega(\mathbf{R}, t) + [\int \phi^*(\mathbf{r}, t)H_r(\mathbf{r}, \mathbf{R})\phi(\mathbf{r}, t)d\mathbf{r}]\Omega(\mathbf{R}, t) \equiv H_n^{\text{eff}}\Omega(\mathbf{R}, t), \quad (4)$$

$$i\hbar\frac{\partial}{\partial t}\phi(\mathbf{r}, t) = [\int \Omega^*(\mathbf{R}, t)H_r(\mathbf{r}, \mathbf{R})\Omega(\mathbf{R}, t)d\mathbf{R}]\phi(\mathbf{r}, t) \quad (5)$$

In conventional Ehrenfest dynamics, a classical approximation is applied to the nuclei, where the nuclear effective Schrödinger equation is approximated by Newton's equation of motion.[10]

However, in this work, we will invoke the constrained minimized energy surface (CMES) approximation, as previously introduced by our group to bridge quantum and classical dynamics.[38] This approximation incorporates NQEs, particularly quantum nuclear delocalization effects, into classical simulations. For clarity in the following derivation, we consider the case of a single quantum nucleus, though generalization to multiple quantum nuclei is straightforward.

In the CMES framework, the nuclear wave function is first expressed in the polar form: $\Omega(\mathbf{R},t) = A(\mathbf{R},t)\exp(iS(\mathbf{R},t)/\hbar)$, where the amplitude part $A(\mathbf{R},t)$ and the phase part $S(\mathbf{R},t)$ are both real functions. Then the nuclear kinetic energy can be expressed as

$$\langle T_{\mathrm{n}} \rangle(t) = \langle A(t) | \hat{T} | A(t) \rangle + \frac{\langle \mathbf{P} \rangle^2(t)}{2M} + \frac{\sigma_{\mathbf{P}}^2(t)}{2M}, \qquad (6)$$

where $\langle \mathbf{P} \rangle$ is the expectation value of the nuclear momentum operator with $\langle \mathbf{P} \rangle(t) = \int d\mathbf{R}\, A^2(\mathbf{R},t)\nabla S(\mathbf{R},t)$, and $\sigma_{\mathbf{P}}^2$ denotes the variance of the momentum field with $\sigma_{\mathbf{P}}^2(t) \equiv \int d\mathbf{R}\, A^2(\mathbf{R},t)[\nabla S(\mathbf{R},t)]^2 - \langle \mathbf{P} \rangle^2$.[38]

Alternatively, according to the definition of $H_{\mathrm{n}}^{\mathrm{eff}}$ in Eq. (4), the nuclear kinetic energy can also be written as

$$\langle T_{\mathrm{n}} \rangle(t) = \langle \Omega | \hat{H}_{\mathrm{n}}^{\mathrm{eff}} | \Omega \rangle(t) - \langle A | \langle \phi | H_r | \phi \rangle | A \rangle(t) \qquad (7)$$

By combining the above two equations and taking the time derivative on both sides, we get

$$\frac{\langle \mathbf{P}\rangle}{M} \cdot \frac{\mathrm{d}\langle \mathbf{P}\rangle}{\mathrm{d}t} = -\langle \frac{\mathrm{d}}{\mathrm{d}t} A \mid \hat{H}_\mathrm{n}^\mathrm{eff} \mid A\rangle - \mathrm{c.c.} - \frac{\mathrm{d}}{\mathrm{d}t}\frac{\sigma_\mathbf{P}^2(t)}{2M}, \quad (8)$$

where "c.c." stands for the complex conjugate of the previous term. Up to this point, all derivations are exact within the mean-field approximation.

Next, we invoke the CMES approximation,[38] where it is assumed that the nuclear state will always adapt to the energy-minimized state for the current effective potential $H_\mathrm{n}^\mathrm{eff}$, while satisfying the constraints on the expectation values of the nuclear position $\langle \mathbf{R}\rangle(t) = \mathbf{R}_t$ and momentum $\langle \mathbf{P}\rangle(t) = \mathbf{P}_t$. Then the Lagrangian for this constrained minimization can be written as

$$\mathcal{L} = \langle \Omega \mid \hat{H}_\mathrm{n}^\mathrm{eff} \mid \Omega\rangle + \mathbf{f} \cdot (\langle \Omega \mid \widehat{\mathbf{R}} \mid \Omega\rangle - \mathbf{R}_t) - \mathbf{v} \cdot (\langle \Omega \mid \widehat{\mathbf{P}} \mid \Omega\rangle - \mathbf{P}_t) - \tilde{E}(\langle \Omega \mid \Omega\rangle - 1), \quad (9)$$

where $\mathbf{f}$ and $\mathbf{v}$ are Lagrange multipliers enforcing the constraints on position and momentum, respectively, and $\tilde{E}$ is the Lagrange multiplier ensuring the wave function normalization. Making the Lagrangian stationary with respect to the variations in $A$ and $S$ (See Supporting Information for details) leads to a uniform momentum field, $\nabla S(\mathbf{R},t) = \mathbf{P}_t$, and an eigenvalue equation for the amplitude part $A$

$$\left[\hat{H}_\mathrm{n}^\mathrm{eff} + \mathbf{f} \cdot (\widehat{\mathbf{R}} - \mathbf{R}_t)\right] \mid A\rangle = E^\mathrm{CMES} \mid A\rangle \quad (10)$$

It can be proved that the eigenvalue $E^\mathrm{CMES}$ satisfies $E^\mathrm{CMES} = \tilde{E} + \frac{\mathbf{P}_t^2}{2M} = \langle A \mid \hat{T} \mid A\rangle + E_r$, meaning that it can be perceived as the sum of the quantum delocalization kinetic energy $\langle A | \hat{T} | A \rangle$ and the instantaneous potential energy $E_r$. Note that the difference in the CMES here versus that in the original theory[38] is that here the quantum evolution of electrons provides an instantaneous potential for the nuclei and the CMES is obtained through constrained minimization in the presence of such a potential. In contrast, in the original theory, the potential used was the static potential with electrons relaxed to the ground state. It should also be noted that the solution $|A\rangle$ depends on both the current classical geometry $\mathbf{R}_t$ and the instantaneous electronic state $\phi$, since $\phi$ enters into the effective Hamiltonian $\hat{H}_n^{\text{eff}}$. Therefore, this amplitude part of the nuclear wave function can be explicitly expressed as $A(\mathbf{R}; \mathbf{R}_t, \phi_t)$.

With the CMES approximation, the uniform momentum field leads to $\sigma_{\mathbf{P}}^2(t) = 0$ and Equation (8) simplifies to $\frac{\langle \mathbf{P} \rangle}{M} \cdot \frac{\mathrm{d}\langle \mathbf{P} \rangle}{\mathrm{d}t} = -\langle \frac{\mathrm{d}}{\mathrm{d}t} A | \hat{H}_n^{\text{eff}} | A \rangle - \text{c.c.}$. Using the facts that $|A\rangle$ satisfies equation (10) and that it depends on the current nuclear geometry $\mathbf{R}_t$ and instantaneous electronic state $\phi$, Equation (8) can be further simplified to

$$\frac{\langle \mathbf{P} \rangle}{M} \cdot \frac{\mathrm{d}\langle \mathbf{P} \rangle}{\mathrm{d}t} = -\frac{\langle \mathbf{P} \rangle}{M} \cdot \nabla_{\mathbf{R}_t} \langle A | \hat{H}_n^{\text{eff}} | A \rangle, \tag{11}$$

with more derivation details provided in the Supporting Information. After the common prefactor $\langle \mathbf{P} \rangle / M$ is dropped following the derivation of the original CMES theory,[38] the final expression is

$$\frac{\mathrm{d}\langle \mathbf{P} \rangle}{\mathrm{d}t} = -\nabla_{\mathbf{R}_t} \langle A | \hat{H}_n^{\text{eff}} | A \rangle. \tag{12}$$

This result means that the nucleus can be perceived as moving on an effective potential energy surface provided by the instantaneous electronic state. For this effective potential energy surface, since $\hat{T}_\text{n}$ is in the effective nuclear Hamiltonian $\hat{H}_\text{n}^\text{eff}$, the zero-point energy and the quantum nuclear delocalization effects are naturally included.

Furthermore, it can be proved that for both analytical case and the case with a fixed nuclear basis set, the gradient of the effective potential energy surfaces is directly related to the Lagrange multiplier with $\nabla_{\mathbf{R}_t} \langle A | \hat{H}_\text{n}^\text{eff} | A \rangle = -\mathbf{f}$. However, if the nuclear basis set moves with the expectation value of the nuclear position, because of Pulay force related terms,[30,31,47] the relationship becomes

$$\nabla_{\mathbf{R}_t} \langle A | (\hat{H}_\text{n}^\text{eff}) | A \rangle = \sum_{ij} c_i \nabla_{\mathbf{R}_t} (\hat{H}_\text{n}^\text{eff})_{ij} c_j, \tag{13}$$

where $c$ is the basis expansion coefficient. The detailed proof for these cases is provided in the Supporting Information.

Equation (12), together with the electronic evolution equation (5), forms the working equation for CNEO-Ehrenfest dynamics. It highly resembles conventional Ehrenfest dynamics with both a Schrödinger-like equation for the electrons and a Newton-like equation for the nucleus. However, the key difference is that here the nuclear evolution is based on the expectation value of the quantum nuclear position, rather than treating nuclei as classical point charges, and the effective potential incorporates quantum nuclear delocalization effects. It can be proved that CNEO-Ehrenfest dynamics, whether analytical or with a basis set, is rigorously energy conserved (See Supporting Information).

We also want to acknowledge the recently developed NEO-Ehrenfest dynamics,[28,29] which propagates both electronic and quantum nuclear wave functions with Schrödinger-like equations, while treating heavy nuclei with Newton-like equations. The key difference between CNEO-Ehrenfest and NEO-Ehrenfest dynamics is that CNEO still essentially treats all electrons quantum mechanically and all nuclei classically, but with an underlying picture that nuclei are quantum-delocalized and possess zero-point energies. In contrast, NEO-Ehrenfest dynamics treats both light nuclei and electrons quantum mechanically, and only heavy nuclei classically.

For practical implementations of CNEO-Ehrenfest dynamics, both diabatic and adiabatic basis sets may be used, which should in principle yield the same results in the complete basis set limit. In the simplest diabatic representation, in which the basis functions $\{\varphi_i(\mathbf{r})\}$ are time-independent, the working equation is

$$i\hbar \frac{\partial}{\partial t} \rho_{ij}(t) = [H_{\mathrm{e}}^{\mathrm{eff}}(t), \rho(t)]_{ij} \tag{14}$$

where $\rho$ is the nuclear density matrix with $\rho_{ij}(t) = c_j^*(t) c_i(t)$. In contrast, for adiabatic representations in which the basis functions $\{\varphi_i(\mathbf{r}; \mathbf{R}_t)\}$ depend parametrically on the current nuclear positions and diagonalize the instantaneous electronic Hamiltonian, $[\int \Omega^*(\mathbf{R}, t) H_r(\mathbf{r}, \mathbf{R}) \Omega(\mathbf{R}, t) d\mathbf{R}] \varphi_i(\mathbf{r}; \mathbf{R}_t) = E_i^{\mathrm{CNEO}}(t) \varphi_i(\mathbf{r}; \mathbf{R}_t)$, the working equation is

$$i\hbar \frac{\partial}{\partial t} \rho_{ij}(t) = \rho_{ij}(t)(E_i^{\mathrm{CNEO}} - E_j^{\mathrm{CNEO}}) + i\hbar [\rho(t), \mathbf{d}]_{ij} \cdot \frac{d\mathbf{R}_t}{dt} \tag{15}$$

with the nonadiabatic coupling vector defined as

$$\mathbf{d}_{ji} = \int \varphi_j^*(\mathbf{r}; \mathbf{R}_t) \nabla_{\mathbf{R}_t} \varphi_i(\mathbf{r}; \mathbf{R}_t) d\mathbf{r} \tag{16}$$

Next, we present the formulation of CNEO surface hopping, which can only be implemented within the adiabatic representation. Similar to conventional surface hopping, the nuclei predominantly evolve on a single potential energy surface but may undergo stochastic hops between surfaces to mimic transitions between electronic states. However, a key difference lies in how the adiabatic energy surfaces are constructed. In conventional surface hopping, these surfaces are obtained by diagonalizing the electronic Hamiltonian with nuclei treated as classical point charges. In contrast, CNEO surface hopping constructs the adiabatic surfaces by diagonalizing the electronic Hamiltonian using a quantum-delocalized nuclear density. Specifically, for a given molecular geometry $\mathbf{R}_t$, we use the nuclear wave function from the ground-state CNEO calculation $A_0(\mathbf{R};\mathbf{R}_t)$, and obtain the adiabatic electronic states at that geometry $\varphi_i(\mathbf{r};\mathbf{R}_t)$ by diagonalizing the electronic Hamiltonian with the fixed ground-state nuclear wave function, i.e.,

$$[\int A_0(\mathbf{R};\mathbf{R}_t)H_r(\mathbf{r},\mathbf{R})A_0(\mathbf{R};\mathbf{R}_t)\mathrm{d}\mathbf{R}]\varphi_i(\mathbf{r};\mathbf{R}_t) = E_i^{\mathrm{CNEO}}\varphi_i(\mathbf{r};\mathbf{R}_t) \qquad (17)$$

This approach utilizes a frozen-nucleus approximation and neglects the nuclear response to electronic excitations. It essentially adopts a Franck-Condon picture for electronic transitions.

In alignment with the conventional FSSH method, the CNEO-FSSH method also defines the hopping probability from state $i$ to state $j$ using the same expression

$$\gamma_{i\to j} = \max\left[0, \frac{2\langle\mathbf{P}\rangle}{M}\cdot\mathrm{Re}(\mathbf{d}_{ij}\rho_{ji}(t))\Delta t \,/\, \rho_{ii}(t)\right] \qquad (18)$$

where $\Delta t$ is the time step.

We employed both the CNEO-Ehrenfest and CNEO-FSSH methods to study the Shin-Metiu model,[46] which describes a one-dimensional system in which a proton and an electron transfer between two fixed ions. The model potential is given by

$$V(r,R) = Ze^2 \left( \frac{1}{|R+\frac{L}{2}|} + \frac{1}{|R-\frac{L}{2}|} - \frac{\text{erf}(\frac{|r+\frac{L}{2}|}{R_{c1}})}{|r+\frac{L}{2}|} - \frac{\text{erf}(\frac{|r-\frac{L}{2}|}{R_{c1}})}{|r-\frac{L}{2}|} \right) - \frac{\text{erf}(\frac{|R-r|}{R_{c2}})}{|R-r|} \quad (19)$$

where $Z$ is the charge of the fixed ions, $e$ is the unit charge, $R_{c1}$ and $R_{c2}$ are adjustable parameters modifying the Coulomb interactions between the electron and fixed ions and the proton, respectively, and $L$ is the distance between the two fixed ions. In the original Shin-Metiu model, the charge number of the fixed ion is set to $Z=1$. However, here to model a PCET process in a real system, we vary it between 0.4 and 0.6 to better describe the mixed covalent and hydrogen bonding nature between the transferring hydrogen atom and the adjacent heavy atoms. We set the $R_c$ parameters to be $R_{c1} = 0.6 \text{ Å}$ and $R_{c2} = 1.6 \text{ Å}$, and $L$ varies between $4.5\text{Å}$ and $5.5\text{Å}$, which gives us a series of models (Models I-V) with different adiabatic barrier heights, energy gaps between the ground and first excited states, and double-well separations (Table 1). These models reflects various realistic electrocatalytic and photosynthesis PCET systems.[48–50] Additionally, Model VI is designed to resemble the phenoxyl/phenol system, with barrier height, energy gap, and well separation closely matching those found in that system.[51]

**Table 1** Model Parameters

| Model | Z | $R_{c1}$ | $R_{c2}$ | L | Barrier (kcal/mol) | Gap (kcal/mol) | Well Distance ($\text{Å}$) |
|---|---|---|---|---|---|---|---|
| I | 0.5 | 0.6 | 1.6 | 5 | 11.8 | 16.0 | 1.55 |
| II | 0.6 | 0.6 | 1.6 | 5 | 12.9 | 10.6 | 1.44 |
| III | 0.4 | 0.6 | 1.6 | 5 | 9.6 | 23.9 | 1.65 |
| IV | 0.5 | 0.6 | 1.6 | 5.5 | 18.4 | 12.1 | 1.98 |
| V | 0.5 | 0.6 | 1.6 | 4.5 | 5.42 | 21.6 | 1.12 |
| VI | 0.45 | 0.1 | 1 | 2.5 | 17.6 | 13.9 | 0.69 |

In our simulations, the proton with a mass of 1836 a.u. is initially placed at the bottom of the left well. In CNEO Ehrenfest dynamics, the starting point is an optimized CNEO ground state in which both the electron and nucleus are localized on the left. The initial momentum is directed to the right, and different simulations are performed with a range of initial momenta, spanning from low to high values. For Ehrenfest dynamics, electronic wave functions are represented using the discrete variable representation (DVR) method[52] and propagated on-the-fly. In CNEO-Ehrenfest dynamics, since the underlying nuclear picture is also quantum, the nuclear wave functions are also treated using DVR with the same choice of DVR Sinc functions as the electronic ones. The quantum dynamics references are also obtained using the DVR method but through the Born-Huang expansion that considers two electronic states. In FSSH and CNEO-FSSH, a total of $2 \times 10^4$ trajectories are simulated for each initial momentum. More computational details are available in the Supporting Information.

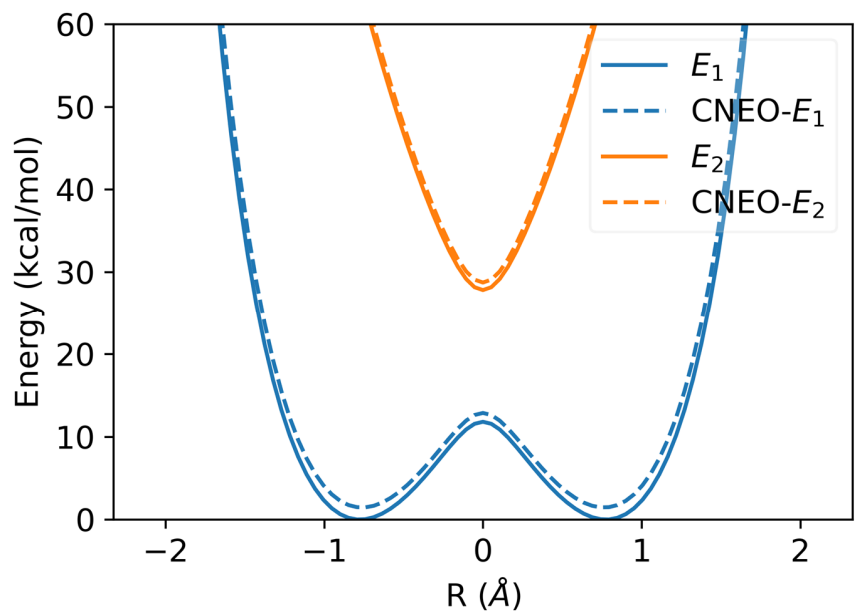

**Figure 1** Ground and first-excited-state PESs and corresponding frozen-nuclear CNEO effective PESs for the Shin-Metiu model with parameter set I. The lowest energy point of the PES is set to zero.

**Figure 1** shows the ground and the first-excited-state PESs, along with the frozen-nuclear CNEO effective potential energy surfaces, for the Shin-Metiu model with parameter set I. The CNEO surfaces lie slightly above the original PESs due to the incorporation of quantum nuclear delocalization energy. Despite this energy shift, the CNEO surfaces preserve similar well separations and a ground-to-excited-state energy gap. Notably, at the transition state, the underlying nuclear wave function is more delocalized, leading to a smaller nuclear energy compared to that in the reactant and product wells. This results in a lower effective reaction barrier on the CNEO ground-state surface. Similar trends are observed for other parameter sets, and their results are provided in **Figure S1**.

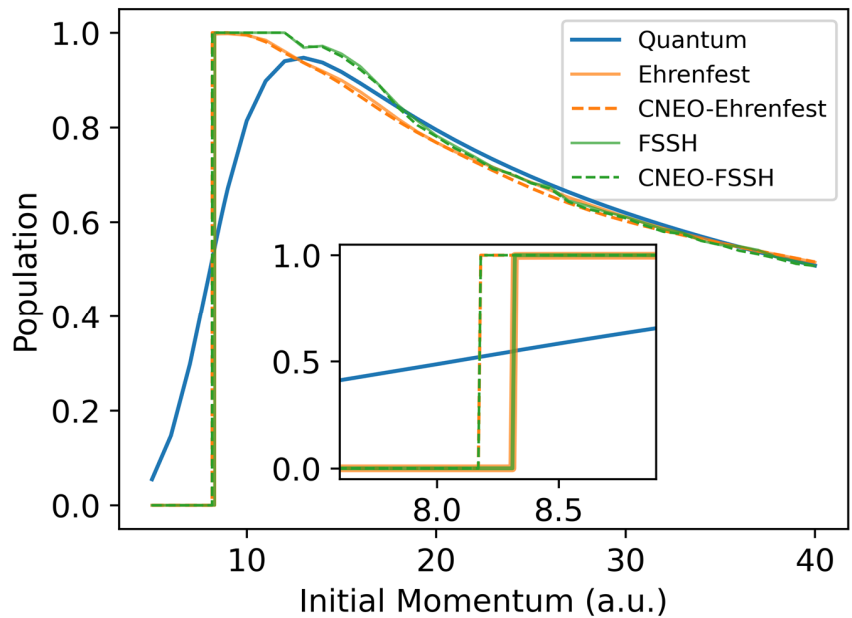

**Figure 2** Transmission probabilities on the lower state as a function of initial momentum for Ehrenfest, CNEO-Ehrenfest, FSSH, and CNEO-FSSH methods, compared to the quantum dynamics reference.

We first investigated the proton transmission probability on the lower state, which is defined as the probability that the system is on the ground adiabatic surface at the first instance when the proton crosses the reaction barrier into the right well. The results for Ehrenfest, CNEO-Ehrenfest, FSSH, and CNEO-FSSH methods as well as the quantum dynamics reference are shown in **Figure 2**. In the high-momentum regime with the initial momentum $k > 20$, all nonadiabatic approaches yield results that closely match the quantum dynamics reference. However, in the low-momentum regime with $k < 10$, the nonadiabatic methods exhibit a sharp step-like transition—showing no transmission below a certain threshold and full transmission above it, whereas the quantum dynamics reference displays a smooth, continuous transmission curve. Notably, the CNEO-based methods exhibit slightly lower momentum thresholds for transmission than their conventional counterparts, which is due to the reduced reaction barrier on the CNEO ground-state surface as discussed earlier.

Given the classical treatment of nuclear motion in both conventional nonadiabatic methods and the CNEO approaches, this step-like feature is expected and has been observed in previous studies.[11,53] However, this behavior makes the direct comparison with the smooth quantum results challenging. To enable a more meaningful analysis, following some treatments by previous studies,[11,53–55] we average the computed transmission probabilities over the initial momentum distribution of the quantum wave packet. This procedure ensures that the classical particles share the same initial momentum distribution as the quantum reference, while keeping their initial positions unchanged. The resulting smoothed transmission curves are shown in **Figure 3**.

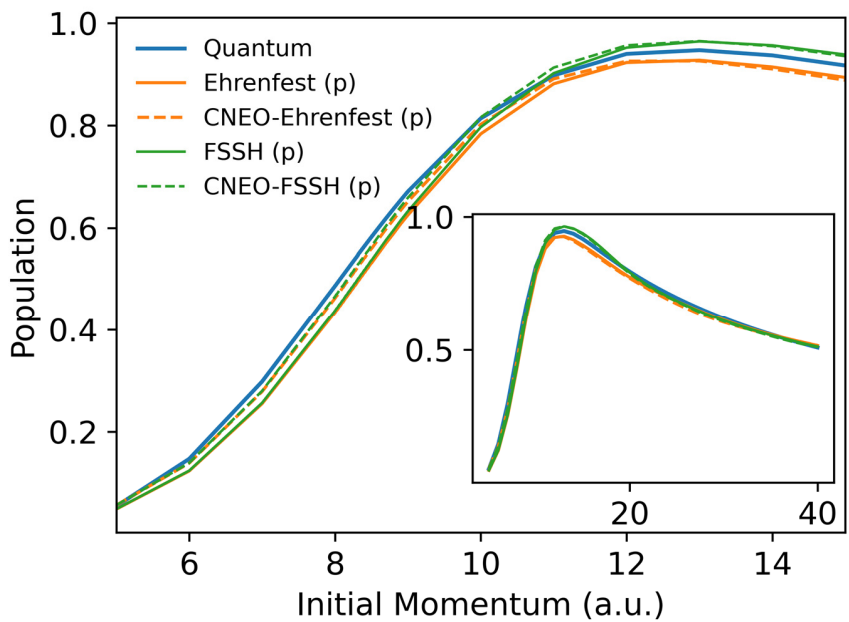

**Figure 3** Transmission probabilities on the lower state for quantum, Ehrenfest, CNEO-Ehrenfest, FSSH, and CNEO-FSSH methods. All nonadiabatic populations are averaged by the initial momentum distribution. The symbol (p) denotes population averaged by the initial momentum distribution.

As shown in **Figure 3**, within the low-momentum regime of 6–10 a.u., the conventional nonadiabatic methods slightly underestimate the transmission probabilities. In contrast, the CNEO-based methods exhibit smaller deviations from the quantum reference, although some underestimation remains. This improved performance is mainly due to the reduced effective barrier through the CNEO treatment. In the mid-momentum regime of 10–20 a.u., both FSSH and CNEO-FSSH tend to overestimate the transmission probabilities, whereas both Ehrenfest and CNEO-Ehrenfest dynamics tend to underestimate them. Although none of the methods perfectly reproduce the quantum results in this regime, the nearly identical performance of CNEO methods and their conventional counterparts suggests that the deviations stem from limitations in the underlying Ehrenfest dynamics or FSSH, rather than from the CNEO treatment. As the initial momentum further increases, the dynamics become more classical in nature. In the high-momentum regime with initial momentum above 20 a.u., all methods yield results that closely match the quantum predictions as shown in the inset of **Figure 3**.

Overall, the CNEO nonadiabatic methods demonstrate improved accuracy in predicting transmission probabilities on the lower state, particularly in the low-momentum regime. This is because nuclear quantum effects are more significant when the momentum is low, and aligns with finding from the previous RPSH study.[27]

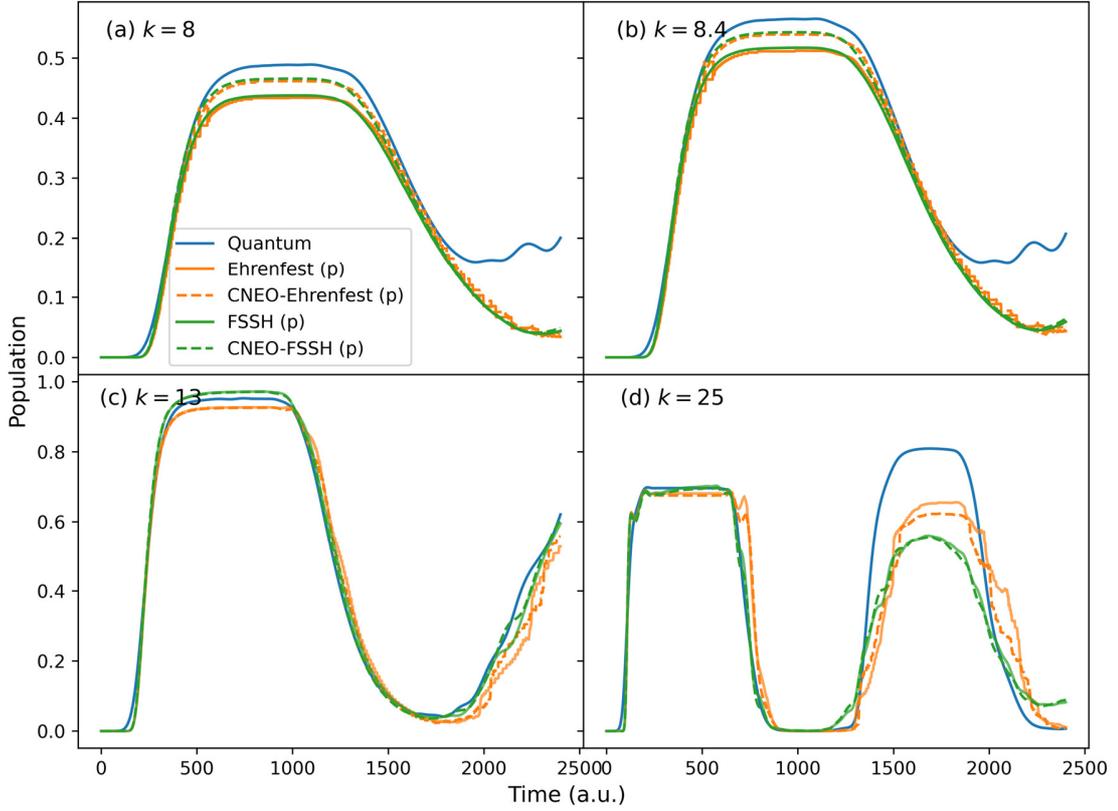

**Figure 4** Time-dependent transmission populations on the lower state for Model I at various initial momenta. Panels a and b correspond to low-momentum cases, panel c represents a mid-momentum case, and panel d shows a high-momentum case.

Next, we dive into the details of the time evolution of the transmission populations on the lower state at different initial momenta with results from four different initial momenta presented in **Figure 4** (see Supporting Information for the computational detail). Panels a and b show results for two momenta in the low-momentum regime. It can be observed that although all methods tend to underestimate the probabilities during the whole dynamics and they all face challenges beyond t > 2000 a.u., CNEO Ehrenfest and CNEO-FSSH overall show the best agreement with the quantum time-dependent populations. Panel c shows results for an initial $k$ in the mid-momentum regime, and we can see the comparable performance between the CNEO methods and their conventional counterparts. As to a high initial momentum case as shown in Panel d, all nonadiabatic methods show good agreement with quantum results in the early stages, although some deviations can be seen during the second barrier crossing stage.

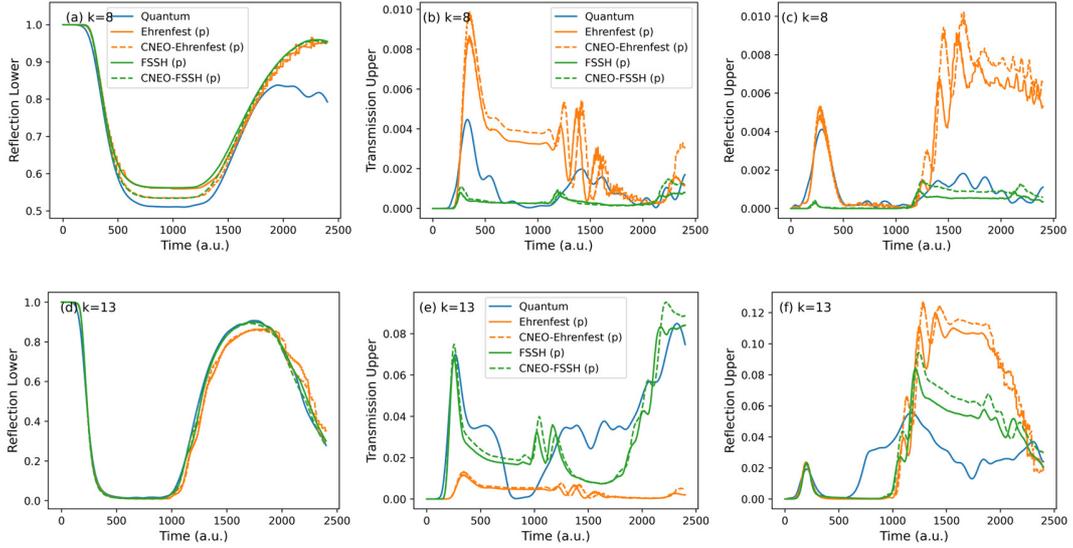

**Figure 5** Time-dependent populations for (a) reflection lower, (b) transmission upper, and (c) reflection upper for Model I at initial momenta of $k=8$ and $k=13$ a.u.

In addition to the transmission probability on the lower state, we also investigate three other populations, namely the reflection lower, transmission upper, and reflection upper populations. **Figure 5** presents the time evolution of these three populations when the initial momentum is $k=8$ and $k=13$ a.u. Results for $k=8.4$ and $k=25$ are provided in the Supporting Information (**Figure S4**). For the reflection populations on the lower state, CNEO methods consistently perform better than or as well as the conventional approaches. For probabilities on the upper surfaces, including both transmission and reflection, all nonadiabatic methods show larger discrepancies. However, the performance in the early stage of the dynamics is usually better.

We further assess the robustness of our method by systematically investigating Models I to V (see Table 1 and **Figure 6**), while reserving Model VI for later discussion. Models I, II, and III explore the effect of varying the ion charge $Z$ to mimic different bonding environments. These three models share similar well separations and ground-state barriers but differ mainly in the ground and excited-state energy gaps. Specifically, Model II has the smallest energy gap, whereas Model III has the largest. In contrast, by comparing Models I, IV, and V, we mainly investigate the impact of the well separation, although we observe that with varying well separations, the energy gaps are also greatly impacted with Model IV having a wider separation and smaller gap while Model V having a narrower separation and a larger gap.

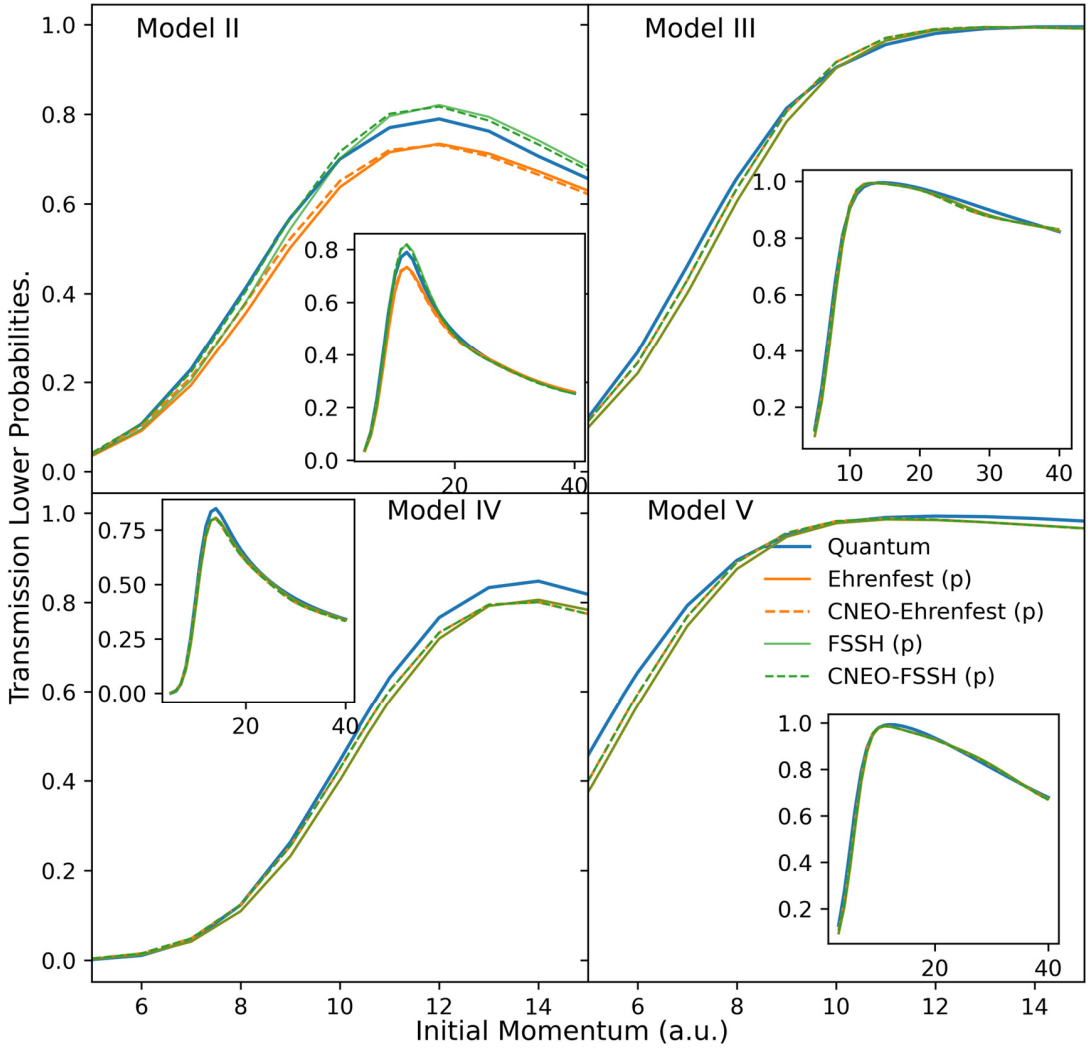

**Figure 6** Transmission probabilities on the lower state for Models II–V as a function of initial momentum.

**Figure 6** shows the transmission probability on the lower state for Model II–V. It can be observed that as the initial momentum increases, all models exhibit an initial rise in transmission probability, followed by a decline. For Models II and IV, this decline starts earlier and occurs more sharply as shown in the insets of the figures. Models II and IV have smaller ground and excited-state energy gaps and therefore a larger number of populations enter the excited state. This further leads to their lower peak transmission probabilities (~0.8). In the low-momentum regime (below 10 a.u. for Model II and 11 a.u. for Model IV), conventional nonadiabatic methods tend to underestimate transmission, whereas the CNEO-based nonadiabatic methods yield higher populations and align more closely with quantum results. Notably, both CNEO-FSSH and CNEO-Ehrenfest perform exceptionally well for Model IV in this regime.

For Models III and V, which feature larger energy gaps between the ground and the excited state, a peak transmission populations close to 1 can be achieved, indicating that for a specific range of momentum, all populations can transmit to the right well without going to the first excited state. In the low-momentum regime, conventional methods again underestimate transmission probabilities, while CNEO-based nonadiabatic methods provide improved—though still slightly underestimated—predictions.

In the Supporting Information **Figure S5–8**, we present the time-dependent transmission population on the lower state at certain initial momenta for Models II–V. Similar to the results observed for Model I, the CNEO methods generally performed better than or at least comparable to the conventional approaches across these different model systems. This consistency further demonstrates the robustness and reliability of the CNEO nonadiabatic methods.

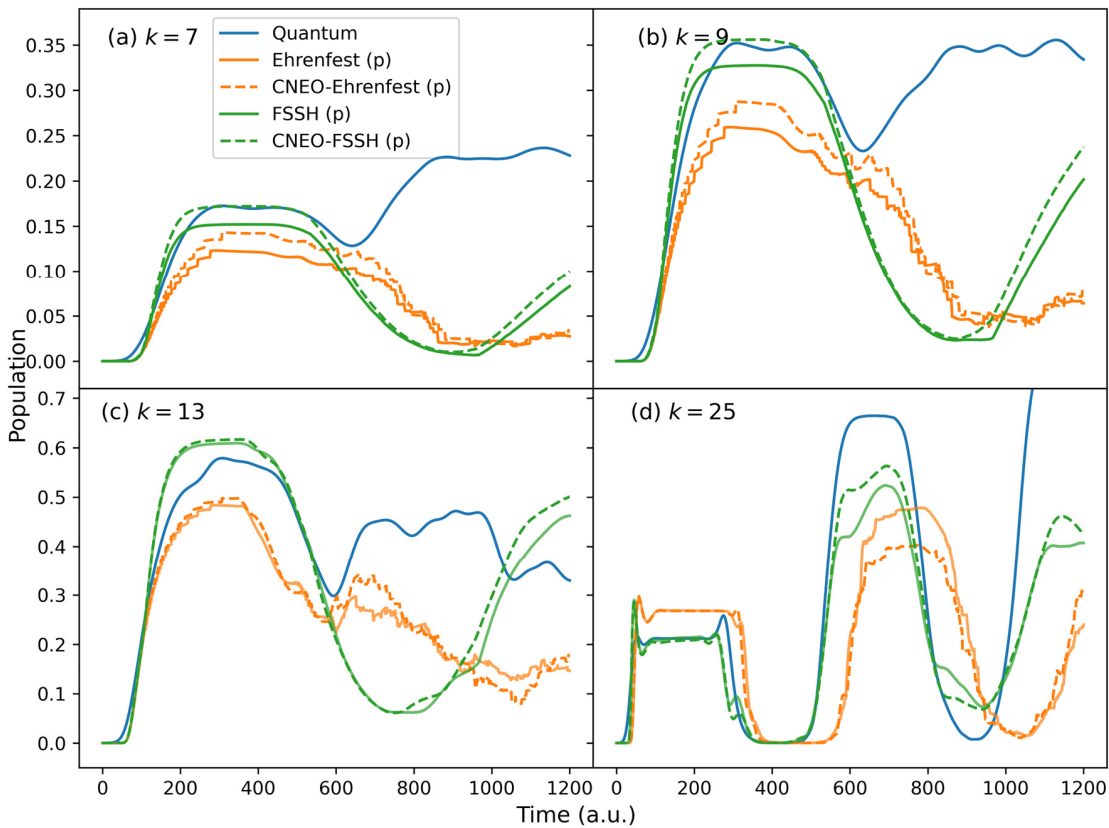

**Figure 7** Time-dependent transmission populations on the lower state for Model VI at various initial momenta.

Finally, we examine model VI, which was specifically designed to mimic the phenoxyl/phenol system. **Figure 7** shows the time-dependent transmission populations on the lower state at four different initial momenta. In the low-momentum regime with $k = 7$ and $k = 9$, the time-dependent population curves of CNEO methods resemble those of conventional methods, but with CNEO-FSSH showing overall the best agreement with the quantum reference. Notably, the quantum population exhibits considerable fluctuations over time, making it difficult to determine an appropriate cutoff time for evaluating transmission probabilities. Therefore, for this model, we only focus on the time-dependent population profiles. In mid- and high-momentum regimes with $k=13$ and $k=25$, NQEs are less significant and all methods are able to predict reasonable population values in a short time frame.

Across all tested models—despite variations in barrier heights, well separations, and ground-to-excited-state energy gaps—the CNEO nonadiabatic methods consistently show improved accuracy over conventional approaches in the low-momentum regime. At mid- and high-momentum regimes, where NQEs are less significant, both CNEO and conventional methods perform similarly. Notably, CNEO-FSSH demonstrates the best overall agreement with quantum results across all models, highlighting its potential as a reliable and broadly applicable method for simulating nonadiabatic processes involving nuclear quantum effects.

In summary, by combining CNEO with Ehrenfest dynamics and FSSH, we have developed two novel approaches for incorporating NQEs into nonadiabatic dynamics. In the CNEO-Ehrenfest method, the quantum nuclear motion is governed by the effective potential energy surface obtained from the instantaneous electronic state. In the CNEO-FSSH method, the adiabatic frozen-nucleus CNEO surfaces with quantum delocalization effects incorporated are employed with a fewest-switches surface hopping algorithm.

Our investigation on the Shin-Metiu model with various parameter choices to mimic real PCET systems demonstrates that the CNEO-based methods significantly improve the prediction of transmission probabilities, particularly in the low-momentum regime. At mid- and high-momentum regimes, where nuclear motion behaves more classically, CNEO and conventional nonadiabatic methods converge toward quantum results, supporting their reliability in these regimes.

Importantly, the utilization of CNEO surfaces introduces only a marginal increase in computational cost compared to using conventional PESs, making it a practical and scalable option for simulating real systems with NQEs. With CNEO excited-state theories currently under development in our group, the CNEO-based nonadiabatic approaches are promising for simulating nonadiabatic processes with accurate descriptions of both NQEs and nonadiabatic couplings. This work lays the foundation for broader applications of CNEO methods to complex nonadiabatic systems such as PCET and photochemical reactions.

**Acknowledgements**

The authors thank Pengfei Huo for helpful discussions. The authors are grateful for the funding support from the National Science Foundation under Grant No. 2238473. The computational resource support from Center for High Throughput Computing at the University of Wisconsin-Madison is acknowledged.[56]

**Supporting Information**

Additional theoretical derivations of CNEO Ehrenfest and CNEO-FSSH Dynamic, the computational details, and supplementary data, including PESs, time-dependent population results and transmission probabilities for various models (PDF).